# Indian policeman's dilemma: A game theoretic model

Kaushik Majumdar, Systems Science and Informatics Unit, Indian Statistical Institute, 8th Mile, Mysore Road, Bangalore 560059, India; E-mail: kmajumdar@isibang.ac.in

**Abstract:** This paper focuses on a one person game called Indian policeman's dilemma (IPD). It represents the internal conflict between emotion and profession of a typical Indian police officer. We have 'split' the game to be played independently by different personality modules of the same player. Each module then appears as an independent individual player of the game. None of the players knows the exact payoff values of any of the others. Only greater than or less than type of inequalities among the payoff values across the players are to be inferred probabilistically. There are two Nash equilibrium (NE) points in this game signifying two completely opposing behavior by the policeman involved. With the help of the probabilistic inequalities probable propensities of the different behaviors have been determined. The model underscores the need for new surveys and data generation. A design of one such survey to measure professionalism of the police personnel has been outlined.

## 1. Introduction

One person games have been discussed by Von Neumann and Morgenstern (Von Neumann & Morgenstern 1953). In a bridge each two player team has been treated as a single player and the two players in the same team as two representatives of that single player (p. 53). This has been termed 'splitting' of a player. More specifically various games of 'patience' and 'solitaire' are considered one person games. *Indian policeman's dilemma*[1] (IPD) is a one person game where the police officer, who happens to be the only player involved, is spilt between two personality tracts namely, 'Professionalism' and 'Emotionality'. It has been assumed that the officer makes various moves either from purely professional consideration or purely emotional consideration. This makes the Professionalism and Emotionality mutually independent. The game has been designed as a two player game played by Professionalism and Emotionality of the same officer.

Like Bridge and Poker there is no chance move involved in the IPD. However that does not mean that everything is certain in this game. The payoff values of the players against the strategies of the opponents are unknown and have to be inferred probabilistically from the events in the environment. In this sense IPD is a game of incomplete information. We will show that in order to reach a solution to the game it will be enough to know about the dominance relationship between certain pairs of payoff values of the two players.

---

[1] Policeman's dilemma refers to the following situation: A devastating fire is engulfing a few vehicles following a road accident. Police and rescuers who have reached the scene realize that not everybody can be saved from being burnt alive despite the best efforts. One motorist who was sure to die in the inferno begged an armed policeman to shoot him so that he could have a less painful death. Euthanasia is illegal, but not killing in such a situation is inhuman. This is called *policeman's dilemma* or PD, which can be modeled as a one person constant sum game with two mutually conflicting strategies.

This game has another significance. It represents an oriental model. While physical, mathematical, chemical, biological, earth and atmospheric sciences are same in the east and in the west, the social sciences tend to be different due to the influence of culture between orient and occident. IPD represents a professional dilemma of the police personnel in many third world countries, particularly India. In more general form it is probably true of any professional human being, when the person has to resolve the conflict between gains through professional commitments and gains through emotional commitments. The utility function here is the maximization of personal gain. Von Neumann and Morgenstern (Von Neumann & Morgenstern 1953) have quantified the gain (p. 86), but in case of the IPD the quantification of personal gain is impossible or extremely difficult to make. Here we haven't tried to quantify the gains at all, but analyzed the game only in terms of gain dominance.

In the next section we describe the game of IPD. In section 3 we give a normal form representation of it. In section 4 we determine its two Nash equilibrium points. In section 5 we probabilistically determine several inequalities among the payoff values, because the actual payoff values are not known. In section 6 we propose a professionalism index for the Indian police personnel. In section 7 we give the complete solution of the game. We conclude the paper with a discussion.

**2. The game**

A wealthy businessman B has violated a law L, which if convicted is punishable by a jail term. B is wealthy and influential enough to involve the area police chief P who is from the federal civil service cadre (B is influential enough to subdue the local police station officers who usually make such arrests). P has two options to book B under L. Subsection 1 of L or L(1) that imposes a maximum prison term of six months and after being arrested the accused is eligible for bail right from the police station. On the other hand L(2) imposes a maximum prison term of two years and it is non-bailable in the sense that the bail can be granted by a court only for which the accused may have to wait till the next morning. In other words, he will have to spend one night in a crowded police lock up with poor amenities. Also under L(2) getting bail will be a little difficult. The longer he stays in custody the more the case will get media attention and it will be increasingly more difficult for him to avoid a sentence. On the other hand, if he can avoid the stigma of being in police custody, taking advantage of his money power and prolonged legal processes, chances are high that he will not be convicted at all, and therefore will never have to go to jail, which will be particularly easier under L(1).

Sensing the arrest is inevitable B tries to fix a deal with P. According to the deal if B is booked under L(1) P's son, who is studying in an expensive private university in the west and having a hard time to arrange the course fee for the next session, will receive a fat scholarship from a trust headed by B. The amount is enough to clear his tuition for the coming session. Of course if B is booked under L(2) P's son receives nothing. He could not qualify for aids from his current university and must take a study loan at high interest to continue. Since both B's son and P's son went to the same (elite) school in the town and have mutually overlapping friend circles this was pretty well known to B.

Only a week ahead of the day of the imminent arrest the ruling party lost the provincial election and the opposition came to power (police is under the control of the provincial government including the federal civil service cadre officers serving in it). In the mean time P's next promotion is due (it happens in time scale). B is a well known fund raiser for the previous ruling party and if he is booked under L(2) P will be offered a glamorous post, which will considerably enhance his status and influence. Alternatively, if B is booked under L(1) P will get a position, although higher than the present one, is widely considered as unattractive. His son's tuition fee for the next session must be cleared within thirty days and the credit situation in his country of residence is positively bad.

**3. The normal form**

From the point of view of 'rational play' B's freedom to choose from among available strategies is rather limited. If P chooses L(1) B is almost certain to get bail soon after the arrest. It is highly unlikely that he would be arrested again until there is a negative verdict by the court and his subsequent bail plea is rejected, chances of which are not too high. So, B does not have to do anything other than following the 'normal procedure'. On the other hand if P chooses L(2) B will come forward with the deal in the hope that P will go back on his decision and will choose L(1) instead. We find it reasonable to model the game as a one person game played by P against his environment in which B simply acts as an agent. The other prominent environmental agents are the prevailing political situation and the cultural tradition (e.g., high level of family commitment). Following is the payoff matrix of the game in its normal form.

|  | | Profession | |
|---|---|---|---|
|  | | L(1) | L(2) |
| Emotion | Fatherhood | $EM_{11}, PF_{11}$ | $EM_{12}, PF_{12}$ |
|  | Promotion | $EM_{21}, PF_{21}$ | $EM_{22}, PF_{22}$ |

P's decision is governed by two competing aspects of his behavior namely, his professionalism (denoted by Profession in the payoff matrix) and emotionality (denoted by Emotion in the payoff matrix). The technical significance of the words profession and emotion is unimportant. In this paper we will assume profession and emotion are two independent entities, not affecting each other.

Profession and Emotion (of P) are the two players of the game, with Profession has two strategies L(1) and L(2), and the two strategies for Emotion are Fatherhood and (a desire of attractive) Promotion. $EM_{ij}$ and $PF_{ij}$ are the payoff values of Emotion and Profession respectively. The values of $EM_{ij}$ and $PF_{ij}$, $i, j \in \{1,2\}$ depend on the environment and we will make probabilistic estimation of some of the inequalities involving them. This will be enough to estimate different decisions that P can make, based on the outcomes of the game.

Clearly the over-arching objective function of the game is achieving satisfaction by P through professionalism and emotionality.

## 4. Nash equilibrium

Any game involving a finite number of players, each with a finite number of pure strategies must have an NE (Nash 1950). Since IPD is a two person game, each with only two pure strategies, it is feasible to find the NE by brute force. From the payoff matrix it is clear that when column player employs L(1) then $EM_{11}$ is the highest payoff for the row player and therefore it is underlined (Equation (1) below). When column player employs L(2), $EM_{22}$ is the highest payoff for the row player and therefore it is underlined (Equation (3)). Similarly, when row player chooses the strategy Fatherhood the highest payoff for the column player to be underlined is $PF_{11}$ (Equation (4)). When row player chooses Promotion the highest payoff for the column player is $PF_{22}$ (Equation (6)). Since $EM_{11}, PF_{11}$ and $EM_{22}, PF_{22}$ pairs have been underlined both the (1,1) and (2,2) cells of the payoff matrix are NE points (Gibbons 1992, p. 9). Now which one will guide the decision of P? Plenty of uncertainties are involved, because we do not know the values of the payoffs. We must resort to probabilistic arguments.

## 5. Probabilistic inferences

In this paper our goal will never be to infer exact values of the entries of the payoff matrix, but to infer with high probability if some greater than or less than type of relationships hold among the values. We will be interested in questions like "What is the probability of $EM_{11} > EM_{21}$ or $p(EM_{11} > EM_{21})$?"

The probability inequalities (1) to (6) below follow from the payoff matrix. (1) holds because the urge of fatherhood requires P to book B under L(1), but the urge for a better promotion in P discourages him to book B under L(1). So $EM_{11}$ is a bigger payoff than $EM_{21}$.

$$p(EM_{11} > EM_{21}) = 1. \tag{1}$$

(2) holds because the commitment of P's fatherhood is more conducive for applying L(1) than L(2) for him on B.

$$p(EM_{11} > EM_{12}) = 1. \tag{2}$$

(3) holds because if L(2) is applied then payoff to desire of promotion will be higher than the commitment of fatherhood on the part of P.

$$p(EM_{22} > EM_{12}) = 1. \tag{3}$$

(4) holds because professionalism of P puts higher value on $PF_{11}$ than $PF_{21}$ when L(1) is to be applied. Note that here unlike in equation (1) there is no question about booking B under L(1). The only issue is to compare between the payoffs when L(1) has been invoked as an inevitable strategy.

$$p(PF_{11} > PF_{21}) = 1. \tag{4}$$

(5) holds because desire of promotion of P will have higher payoff value when L(2) is applied on B rather than L(1).

$$p(EM_{22} > EM_{21}) = 1. \tag{5}$$

(6) holds because professionalism of P puts higher value on $PF_{22}$ than $PF_{12}$ when L(2) is to be applied.
$$p(PF_{22} > PF_{12}) = 1. \tag{6}$$

From section 4 we know that $EM_{11}, PF_{11}$ and $EM_{22}, PF_{22}$ are NE representing two different behaviors by P. $EM_{11}, PF_{11}$ is for booking B under L(1) and accepting the scholarship for P's son. $EM_{22}, PF_{22}$ is for booking B under L(2) and accept the desired promotion. In each of the next two sections (4 and 5) we will establish one probabilistic inequality. These inequalities will determine the probability of the NE to be chosen by P. In section 7 we will find the complete solution of the game.

Unlike Harsanyi (Harsanyi 1967) here we will be dealing with objective probability, because subjective probability makes sense only with respect to P, but not with respect to the players Profession and Emotion in our game. P's judgment (manifested through Profession and Emotion) will be based on his learning from his environment (both social and natural) in the course of his life. The comparison $EM_{11} > EM_{22}$ in his mind will be an objective Bayesian probability conditional to B's offer and government's offers.

Let $\Omega$ be the space of all events occurring in P's environment, which incorporates the events $e_1$ = B will offer a scholarship to P's son, $e_2$ = The state government will offer P the promotion 1 (desired by P), $e_3$ = The state government will offer P the promotion 2 (not so much desired by P). It is clear $e_i \cap e_j = \varphi, i, j \in \{1,2,3\}, i \neq j$, and since all these three events make up the event space $\Omega = \cup_i e_i$, $i \in \{1,2,3\}$. Probability measures of the events, under the assumption of independent identical distribution (i.i.d), are as following

$$p(e_1) = p(e_2) = p(e_3) = \frac{1}{3}. \tag{7}$$

We are interested in the probability of the events $em_{12} = \{EM_{11} > EM_{22}\}$ and $pf_{21} = \{PF_{22} > PF_{11}\}$ i.e., $p(em_{12})$ and $p(pf_{21})$ respectively. In this section we will derive $p(em_{12})$ and we will take up $p(pf_{21})$ in the next.

**Theorem 1:** Under the assumption of equation (7) if $p(e_2 | em_{12}) = p(e_2)$ and $p(e_3 | em_{12}) = p(e_3)$ then $p(em_{12} | e_1) = p(em_{12})$. In other words, if the government's (or of the political party in power) decision of promotion for P is not influenced by the concern for his emotional tendencies then P's bribe payoff considerations remain unaffected by an offer of bribe (a more realistic interpretation appears after the proof).

**Proof:** Let us assume $p(e_2 | em_{12}) = p(e_2)$ and $p(e_3 | em_{12}) = p(e_3)$ both hold good.
By Bayes' rule we can write

$$p(e_1 | em_{12}) = \frac{p(e_1) p(em_{12} | e_1)}{\sum_{k=1}^{3} p(e_k) p(em_{12} | e_k)}. \tag{8}$$

Let

$$p(e_1 | em_{12}) = \alpha, \tag{9}$$

where $0 \leq \alpha \leq 1$. By Bayes' rule we have

$$p(em_{12} | e_2) = \frac{p(em_{12}) p(e_2 | em_{12})}{p(e_2)}, \tag{10}$$

in which $p(e_2 | em_{12}) = p(e_2)$. So by (10) it implies $p(em_{12} | e_2) = p(em_{12})$. Similarly from $p(e_3 | em_{12}) = p(e_3)$ we get $p(em_{12} | e_3) = p(em_{12})$.
From (7), (8), (9), (10) we get

$$p(em_{12} | e_1) = \frac{2\alpha}{1-\alpha} p(em_{12}). \tag{11}$$

Or, $\frac{p(em_{12} \cap e_1)}{p(e_1)} = \frac{2\alpha}{1-\alpha} p(em_{12}) \Rightarrow p(e_1 | em_{12}) = \frac{2}{3}\frac{\alpha}{1-\alpha}$. Since $\alpha \neq 0$, by (9) we derive

$$\alpha = \frac{1}{3}. \qquad (12)$$

Substituting (12) in (9) we have

$$p(e_1 | em_{12}) = \frac{1}{3} = p(e_1), \qquad (13)$$

which implies $p(em_{12} | e_1) = p(e_1)$. □

In some countries attractive promotions are often offered as a reward for loyalty rather than professional excellence. If this stops commitment towards professionalism will increase and therefore tendency to accept bribe will decrease. If the assertion of Theorem 1 is viewed this way it will look more convincing than it appears initially. Theorem 1 also supports Harsanyi's observation, "One of the great intellectual achievements of the twentieth century is the Bayesian theory of rational behavior under risk and uncertainty." (Harsanyi, 1978).

Since independence of $em_{12}$ and $e_1$ is not a practical proposition we have to admit that $em_{12}$ is dependent on $e_i$, $i \in \{1, 2, 3\}$. By (7) it is easy to establish

$$3^{-1}(p(em_{12} | e_1) + p(em_{12} | e_2) + p(em_{12} | e_3)) = p(em_{12}). \qquad (14)$$

The cumulative amount of deviation from the ethical and the legal standards that might be termed as *corruption* is $p(em_{12} | e_1) + p(em_{12} | e_2) + p(em_{12} | e_3)$ (from section 4 it is clear that $p(em_{12})$ and $p(pf_{21})$ are the quantities which will determine P's decision, out of which $p(em_{12})$ signifies the propensity to accept B's offer and therefore signifies corruption in P's mentality) to which we assign values according to the following formula

$$p(em_{12} | e_1) + p(em_{12} | e_2) + p(em_{12} | e_3) = r\left(1 - \frac{1}{\sqrt{2\pi.10}} \int_{\sqrt{C}}^{\infty} e^{-\frac{x^2}{2.10}} dx\right), \qquad (15)$$

where $r = p(EM_{22})$ ($0 < r < 1$) is the probability of choosing $EM_{22}$, $C$ is the latest available Corruption Perceptions Index (CPI) prepared by the Transparency International (TI) ($1 < C < 10$, 180 countries listed so far with the lowest score 1.1, the higher is $C$ the less corrupt the nation is). 10 signifies the 10 point scoring scale of CPI. In 2009 India had $C = 3.4$ (CPI 2009). By (14) and (15) we get

$$p(em_{12}) = \frac{r}{3}\left(1 - \frac{1}{\sqrt{2\pi \cdot 10}} \int_{\sqrt{C}}^{\infty} e^{-\frac{x^2}{2 \cdot 10}} dx\right). \qquad (16)$$

For India $p(em_{12}) = 0.3090\,r$. It is clear that $r$ is an important quantity, in other words, the payoff value assigned to $EM_{22}$ is important. $EM_{22}$ is of course person specific and culture specific. If $EM_{22}$ is taken to be larger than the usual value (the so called mean value of an average population) then $r$ will be small and so $p(em_{12})$. Let us state our second important result in the form of

**Theorem 2:** There is less than 31% chance that P will be motivated to accept B's offer.

**Proof:** $p(em_{12}) = 0.3090\,r < 0.31$, because $0 < r < 1$. □

Now we will be estimating $p(pf_{21})$. It is not clear from the description of the game whether B's offence is serious enough to warranty L(2). If there is strong circumstantial evidence to book him under L(2) then obviously $p(PF_{12} > PF_{11}) = 1$. If on the other hand B's prima-facie offence is not strong enough to sustain L(2) against him then $p(PF_{11} > PF_{12}) > \frac{1}{2}$. So we have the following two cases.

Case 1: $p(PF_{12} > PF_{11}) = 1$

From (6) we have $p(PF_{22} > PF_{12}) = 1$. $p(pf_{21}) = p((PF_{22} > PF_{12}) \cap (PF_{12} > PF_{11}))$. $p((PF_{22} > PF_{12}) \cap (PF_{12} > PF_{11})) = p(PF_{22} > PF_{12}) \times p(PF_{12} > PF_{11})$, because $PF_{22} > PF_{12}$ and $PF_{12} > PF_{11}$ are mutually independent. So

$$p(pf_{21}) = 1. \qquad (17)$$

Case 2: $p(PF_{11} > PF_{12}) > \frac{1}{2}$

The question is "What is the probability $p(pf_{21})$?" Notice that here we do not have CPI type of data provided by the TI. To the best of knowledge of the author no survey of professionalism of the Indian police personnel ever existed. Therefore a professionalism survey for the policemen has been outlined in the next section.

**6. Professionalism index**

Professionalism has been quantified by professional-index or p-index by Hoy (McMahon & Hoy 2009, Hoy 2009). This p-index was originally designed to measure professionalism (in terms of a score) of the American school teachers. Many of the 8 attributes on which professionalism has been measured will be applicable for measuring professionalism of the police personnel also. Here we have adopted the original p-index of Hoy in a modified manner to test the professionalism of the Indian police officers. The 7 statements (to be confirmed or denied by the respondent by one of six different choices) that make up the survey questionnaire are as following: (1) I am serious about intervening whenever there is a breach of law in my sight even during the off duty hours although it does not bring me any additional incentive from my department. (2) My colleagues do not give me a lot of credit for being an effective policeperson. (3) I welcome critical opinion on my performance. (4) I do not keep myself updated about the latest interpretations of the Indian Penal Code, (Indian) Criminal Procedure Code, Indian Evidence Act and Indian Police Act, because this is the public prosecutor's job. (5) Professional development is usually a waste of time. (6) I find departmental meetings reviewing the law and order situations from time to time useless, because they do not give me more power to deal with the violation of laws. (7) Police personnel have a responsibility to participate in the public relations exercises. The six choices of answer are (a) strongly disagree, (b) disagree, (c) somewhat disagree, (d) somewhat agree, (e) agree, (f) strongly agree. The scoring protocol of Hoy (Hoy 2009) and McMahon & Hoy (McMahon & Hoy 2009) has been followed. This score has been normalized to a 10 point scale to produce the p-index for our current survey. Normalization keeps the scale of CPI and p-index the same. The higher is the score the more is the professionalism.

Till the time of submitting this manuscript we could not gather response of a single police officer to the survey. For multiple responses the average p-index is to be taken as the p-index of the sample. If the sample is representative and large enough, national p-index for the police can be statistically inferred. If it can be done for each country in the world then the countries can be ranked according to the professionalism of their police department. It may be worth observing that the notion of modern professionalism is very much based on capitalistic value systems. The higher the capitalism in a society the more deep rooted will be the capitalistic value system, and stronger will be the sense of professionalism among the individuals of that society. In such a society police will invariably be more professional than in a relatively feudal society. Capitalistically India is backward than the West European, North American and some East Asian countries. We quite arbitrarily assign the international rank 30 to India according to the national police p-index, which we take as 6.5 (the higher the rank and the p-index the more professional the police of that country is).

### 7. Complete solution

In view of inequalities (1) to (6) it is clear that the NE and therefore the solution of the IPD game, which will decide the course of decision of P, completely hinges on precisely two events namely, $em_{12}$ and $pf_{21}$. So $p(em_{12})$ and $p(pf_{21})$ will guide the decision of P. We have an upper bound on $p(em_{12})$ due to Theorem 2. In view of the p-index survey we have a CPI type index available to us for measuring professionalism of the police personnel. So we are in a position to adopt a similar formulation for $p(pf_{21})$ also. Accordingly, following (15) we write

$$p(pf_{21} \mid e_1) + p(pf_{21} \mid e_2) + p(pf_{21} \mid e_3) = s\left(1 - \frac{1}{\sqrt{2\pi}.10} \int_{\sqrt{Q}}^{\infty} e^{-\frac{x^2}{2.10}} dx\right), \qquad (18)$$

where $s = p(PF_{11})$ ($0 < s < 1$), $Q$ is the p-index for the Indian policepersons (taken to be 6.5) and 10 for the 10 point scoring scale adopted for the p-index. Following (16) we get

$$p(pf_{21}) = \frac{s}{3}\left(1 - \frac{1}{\sqrt{2\pi}.10} \int_{\sqrt{Q}}^{\infty} e^{-\frac{x^2}{2.10}} dx\right). \qquad (19)$$

According to the above data India $p(pf_{21}) = 0.2999s$ (unlike $p(em_{12})$, $p(pf_{21})$ has been calculated on fictitious data, however exactly the same calculation will carry over to real data when they will be available).

**Theorem 3:** When B's prima-facie offence is not strong enough to sustain L(2) against him then, there will be less than 30% chance that B will be booked under L(2).

**Proof:** $p(pf_{21}) = 0.2999s < 0.30$, because $0 < s < 1$. □

By Theorem 2 we have $p(em_{12}) < 0.31$. So $p(EM_{22} > EM_{11}) > 0.69$ and by Case 1 of section 5, $p(pf_{21}) = 1$ (equation (17)). Hence the probability that P will be guided by the NE point in cell (2,2) of the payoff matrix is $> 0.69.1 = 0.69$. That is, if there is strong circumstantial evidence to book B under L(2), there will be greater than 69% chance that he will be booked under L(2). From Theorem 2 it is clear that lower the value of $r$ the higher the chance is for P to book B under L(2) in the situation of Case 1 in section 5. Let us recall that $r = p(EM_{22})$. $r$ is the probability density function of the values that can be assigned to $EM_{22}$. If $EM_{22}$ is kept high, much above average, then obviously $r$ will be small and the probability that P will choose the cell (2,2) as the point to behave in accordance with will be $\gg 0.69$. In other words if P is over jealous for promotion then B has little chance to escape L(2) under the situation of Case 1, which is quite reasonable.

In the situation of Case 2 in section 5, in place of equation (17) Theorem 3 will govern the value of $p(pf_{21})$. So $p(pf_{21}) \leq 0.2999$. By Theorem 2, $1 - p(em_{12}) \geq 0.691$. The probability that P's action will be guided by the NE point, cell (2,2), in the payoff matrix is $p(pf_{21})(1 - p(em_{12}))$. A lot will depend on $r$ and $s$, but $p(pf_{21})(1 - p(em_{12}))$ cannot be greater than $0.2999$. So even if P is over jealous for promotion ($r \to 0$) and his professionalism is very much average ($s \to 1$) the chances of B's booked under L(2) is not more than $0.2999$.

The probability that P's action will be guided by the NE point, cell (1,1), in the payoff matrix is $p(em_{12})(1 - p(pf_{21}))$. By Theorem 2 $p(em_{12}) \leq 0.3090$ and by Theorem 3 $1 - p(pf_{21}) \geq 0.7001$. That is, the probability of P's accepting B's offer is never more than $0.3090$.

We can summarize – if there is a strong prima facie evidence to book B under L(2) then there is more than 69% chance that he will be booked under L(2). His effort to bribe P has less than 31% chance to succeed. When prima facie evidence against B to book him under L(2) is not strong enough he has 29.99% or less chance of being booked under L(2). In this particular situation the bribing effort is likely to be more successful. In other words if P is too politically motivated to appease the ruling party a suitable bribe can upset his effort. Corruption progenies corruption. This again supports the assertion of Theorem 1.

**Discussions**

In this paper we have presented a game theoretic model of how a police officer is resolving the conflict between his professional commitment and personal commitment. This is a one person game, that has been split into a two person game of inner-conflicts of the same person. This clearly shows a way to how a one person game can be formulated as an n-person game, which was originally been suggested by Von Neumann and Morgenstern (Von Neumann & Morgenstern 1953).

Unlike the prisoner's dilemma, the IPD does not have clearly determined payoff values for all the strategies. In case of the IPD determining the exact payoff values is difficult (sometimes may even be impossible). Here we have seen that some greater than or less than type of relationships can be inferred probabilistically and that is good enough to predict the behavior of the player of the one person game.

IPD has been formulated in the backdrop of the Indian society, where the cultures are different from the west. Equation (7) is a simplistic but very important assumption. In a more realistic situation the event space will be larger with unequal probabilities for the individual members. Particularly for unequal probability measure for different events (the event space makes up the environment of the player) almost all the subsequent calculations will become very complicated. One way to tackle this might be Bayesian inferences based on suitable priors. We have already exhibited the power of Bayesian inference in Theorem 1. But it is based on the over simplistic assumption of equation (7).

Equations (15) and (18) need to be built up based on proper surveys. Only a survey of appropriate magnitude can indicate which probability distribution should be taken. In absence of any such survey we have chosen the normal distribution. This however has some significant shortcomings. Under (7) and (15) an average officer from a country with high CPI score (which means less corrupt) is only marginally less inclined to personal gratification than an average officer from a country with low CPI score (means more corrupt). Similarly under (7) and (18) an officer with high p-index in the situation of IPD will only have marginally more professional commitment than an officer with low p-index. This underscores the need to reformulate (7), (15) and (18) with adequate practical data, which is lacking at present.

In this paper in order to get appropriate data for solving the IPD a professionalism survey for the policemen has been proposed. The survey has been designed on the line of a professionalism index proposed for the school teachers in the USA (Hoy 2009). This has been suitably modified to propose a professionalism index for the Indian police personnel. However the survey is yet to be conducted on a sufficient number of police personnel.

**References**

Corruption Perceptions Index (CPI) 2009, Transparency International (TI), http://www.transparency.org/policy_research/surveys_indices/cpi/2009/cpi_2009_table

Gibbons, R., 1992. A Primer in Game Theory, Pearson Education Limited, England.

Harsanyi, J. C., 1967. Games with incomplete information played by "Bayesian players", I-III, Part – I. The basic model, Management Science 14(3): 159 – 182.

Harsanyi, J. C., 1978. Bayesian decision theory and utilitarian ethics, The American Economic Review 68(2): 223-228.

Hoy, W. K., 2009. Professionalism Index (P-Index), available at http://www.waynekhoy.com/professionalism_index.html

McMahon, E. and Hoy, W. K., 2009. Professionalism in teaching: Toward a structural theory of professionalism, In: Studies in School Improvement, Hoy, W. K. and DiPaola, M., (Eds.), Information Age Publishing Inc., 205 – 230.

Nash, J. F., 1950. Equilibrium points in n-person games, Proc. Nat. Acad. Sc. (USA) 36: 48-49.

Von Neumann, J. and Morgenstern, O., 1953. Theory of Games and Economic Behavior, 3rd ed., Princeton University Press, Princeton.